# ORBITAL EVOLUTION OF 4179 TOUTATIS

SURYADI SIREGAR [#], ENDANG SOEGIARTINI

*Astronomy Research Division and Bosscha Observatory, Faculty of Mathematics and Natural Sciences, Bandung Institute of Technology, Ganesha 10, Bandung, Indonesia 40132.*

[#] *Email: suryadi@as.itb.ac.id*

**Abstract**. Asteroid 1934 CT;1989 AC, well known as 4179 Toutatis, is an Apollo and Mars-crosser asteroid with a chaotic orbit produced by a 3:1 resonance with Jupiter and a 1:4 resonance with the Earth, and frequent close approaches to the Earth. It is listed as a potential hazardous object (PHA). The aim of this study is to investigate the possibility of 4179 Toutatis to be ejected from the Solar System. This paper presents an orbital evolution of 4179 Toutatis in time interval of ~300 kyr. Investigation of its orbital evolution is conducted by using the Mercury subroutine package, where the gravitational perturbations of eight major planets in the Solar System are considered. Over very short time scales (~300 kyr) relative to the Solar System life time (~10 Gyr), the asteroid 4179 Toutatis gave an example of chaotic motion that can cause asteroid to move outward and may be followed by escaping from the Solar System.

## 1. INTRODUCTION

The asteroid 4179 Toutatis was first sighted on February 10, 1934 as object 1934 CT, and then promptly lost. It remained a lost asteroid for several decades until it was recovered on January 4, 1989 by Christian Pollas. According to (Williams 2004) it is a double object probably in contact, estimated one 2.5 km and one 1.5 km diameter having mass $5 \times 10^{13}$ kg, rotation period 130 hrs. The orbital period is 4.02 yrs and spectral class S. The last close approach to Earth about 0.050 au happened on November 9, 2008. Asteroids with this characteristic well known as Near Earth Asteroid (abbreviated NEA). These asteroids are classified into four classes: Atira, Aten, Apollo, and Amor (AAAA) types. At time being more than 8000 orbital elements of NEA already determined. Toutatis is prototypical of Apollo with perihelion distance, $q < 1.0167$ au and semi major axis, $a > 1.0$ au the orbit overlap with the Earth in the region of perihelion. NEAs are widely believed to be continuously injecting into Earth-approaching orbits from main-belt asteroids as consequence of energetic inter-asteroidal collisions. The orbit with the perihelion distance $q = 0.94$ au and a small inclination $i = 0°.45$ hence Toutatis had good conditions to approach closely the Earth.

By definition all of asteroids with a minimum orbit intersection distance (MOID) of 0.05 au or less and an absolute magnitude (H) of 22.0 or less are considered as Potentially Hazardous Asteroids (PHAs). This "potential" to make close Earth approaches does not mean a PHA will impact the Earth. It only means there is a possibility for such a threat. By monitoring these PHAs and updating their orbital elements, we can better predict the close-approach statistics and thus their Earth-impact threat.

The hazards of asteroid or comet collisions are direct ones and indirect (afterward) ones. The direct hazards are a huge earthquake, a tsunami and an urban and wild fire. The indirect hazards are the large temperature drop, acid rain fall, and depletion of ozone layer. These direct hazards are local a short term, but the indirect hazards will be worldwide and long duration. The worldwide temperature drop that has serious effect on photosynthesis is most important to a crop production and in turn to surviving of human being.

## 2. ORBITAL EVOLUTION SCHEME

Let a system of *n*-bodies consist of point masses $M_i$ at $\mathbf{r_i}$ where i=1,2,…*n* . Consider the equation of motion of a body of negligible mass, an asteroid in Solar System. If origin is taken at the center of the Sun and **r** is the position vector of the asteroid, we have the acceleration vector of asteroid motion given by:

$$\ddot{r} = -GM_\odot \frac{r}{|r|^3} - \sum_{i=1}^{n} GM_i \left\{ \frac{r - r_i}{|r - r_i|^3} + \frac{r_i}{|r_i|^3} \right\} \quad (1)$$

where *G*- is gravitational constant, $M_\odot$ is mass of the Sun, and $M_i$ is mass of the ith planet. The first term represents acceleration force due to the Sun, the second and third one is acceleration force by planets to asteroid, and planet to the Sun (Aarseth 2003). Dynamical evolution of the asteroid 4179 Toutatis was represent by inspecting positions and velocities of the asteroid and all planets. This was done by employing numerical integration of general Cartesian-N-Body Newtonian equation of motion along the time-evolution. The asteroid is assumed to be mass-less. By taking a time step of 1/1000 yr~1/240 of Mercury's period, and considering the perturbation of all planets in the Solar System, we deduced that relative accuracies of the integration attain maximum at an order of $10^{-6}$ in total energy and angular momentum which are appropriate enough for a conservative system. The MERCURY package used in this work is designed to integrate a set of mutually gravitationally interacting bodies together with a group of test particle under the gravitational influence of the massive bodies but does not affect each other. The orbit for Toutatis is one of the best determined of any asteroid and these is no chance that this object will collide with the Earth during this encounter or any other encounter for at least 500 years (Yeomans & Chodas 2004). This section presented simulation done for 4179 Toutatis.

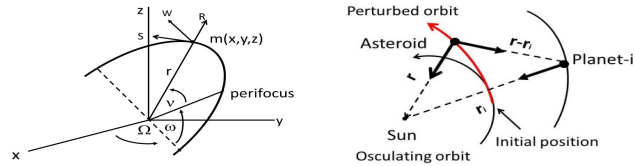

Figure 1. Perturbing planet changes the element orbital of asteroid as function of time *t*.

In figure 1, perturbing radial force R and perturbing transversal force S work on the orbital planes and W is perturbing force that perpendicular on the orbital plane. The alteration of vector radii r will influences the orbital elements.

## 3. RESULTS

The final eccentricity, semi major axis, perihelion distance, inclination, ascending node, argument of perihelion, period in year, daily motion, aphelion and heliocentric distances are presented in Table 1.

Table 1. Orbital elements of 4179 Toutatis now (JDE2456132.5) and 300 kyr later ( exact number t~314053 yr).

| No | Element | Now | Later |
|---|---|---|---|
| 1 | $e$ | 0.6294 | 0.9043 |
| 2 | $a$ (au) | 2.5292 | 67.5945 |
| 3 | $q$ (au) | 0.9372 | 6.4658 |
| 4 | $i$ (deg) | 0.4460 | 16.8943 |
| 5 | $\Omega$ (deg) | 278.5613 | 69.4557 |
| 6 | $\omega$ (deg) | 124.5087 | 174.4445 |
| 7 | $P$ (year) | 4.0224 | 154419.9849 |
| 9 | $Q$ (au) | 4.1212 | 128.7231 |

Over 300 kyr the mass of the Sun is relatively constant, and also the orbit of planets both outer and inner are essentially not affected by the Solar System gravitation. The constancy of semi major axes of planets was demonstrated in previous studies, for example see Siregar (2010). These situations are completely different for 4179 Toutatis, whose orbital eccentricity and semi major axis changes rapidly. We found that the orbit is not stable and chaotic as can be seen in Figures 2.

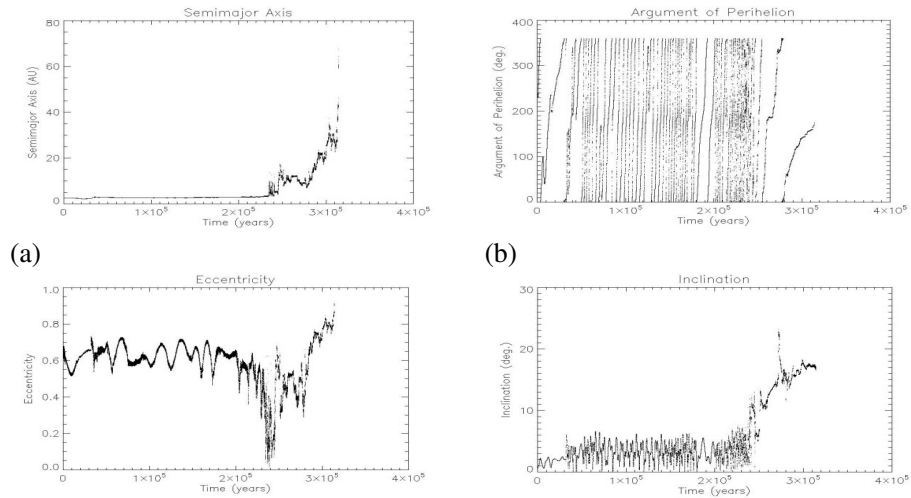

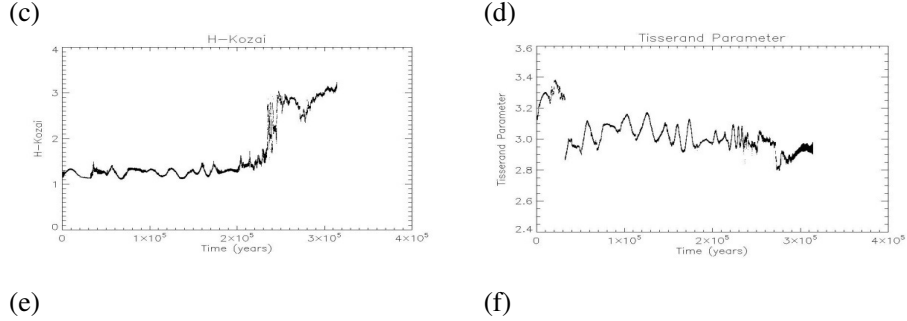

(e)   (f)

Figure 2. Evolution of semimajor axis, (*a*) in au, argument of perihelion (ω) in degree, eccentricity (*e*) inclination (*i*) in degree, the Kozai resonance (H), and Tisserand invariant. Integration time-step is $10^{-3}$ year by including all of planet's perturbation. A transfer of orbital shape from elliptic to parabolic exists at time interval *t* ~314053 year. Integration is stopped at *t* ~314053 year

After 200 kyr the inclination of Toutatis shows the increasing so the possibility the asteroid orbit crossing the Earth's orbit tends to zero. The data of minimum distance to the Earth is given by http://neo.nasa.orbit tell us that Toutatis should be considered as PHA in the periode 1992, 1996, 2004, 2008, 2012, and 2069.

Table 2. Approaches of 4179 Toutatis to the Earth closer than 0.4 au in the periode 1906-2074, minimum distance ρ is given in a.u (see also; Sitarski 1998)

| Date | ρ | Date | ρ | Date | ρ |
|---|---|---|---|---|---|
| 1906-Oct-02 .61 | 0.155 | 1988-Dec-26.00 | 0.117 | 2008-Nov-09.52 | 0.050 |
| 1933-Dec-09.71 | 0.106 | 1992-Dec-08.23 | 0.024 | 2012-Dec-12.28 | 0.046 |
| 1937-Aug-23.99 | 0.191 | 1996-Nov-29.95 | 0.035 | 2016-Dec-29.59 | 0.251 |
| 1981-Jan-12.99 | 0.377 | 2000-Oct-31.19 | 0.074 | 2065-Sep-05.27 | 0.364 |
| 1985-Jan-05.71 | 0.282 | 2004-Sep-29.57 | 0.010 | 2069-Nov-05.66 | 0.020 |

## 4. DISCUSSION AND CONCLUSION

After 300 kyr the Tisserand invariant shows relatively stable at $T<3$ this is contrary with our knowledge about the Tisserand criteria where in general the asteroid has $T \geq 3$ meanwhile comet has $T<3$. As comparison previous study by Siregar (2005) shows that the orbit of 4179 Toutatis tend to T=3.0297 there exist the migration from the elliptic orbit to the parabolic orbit after 300 kyr from now. The escape velocity of 4179Toutatis at aphelion is 3.7 km s$^{-1}$ while its orbital velocity is 0.8 km s$^{-1}$. The impulse of Δ =2.9 kms$^{-1}$ are able to eject asteroid from its orbit. This suggests that the asteroid cannot sustain its orbit in the Solar System. Will the 4179 Toutatis escapes from our Solar System? Asteroid showed no serious threat to the Earth's, smallest MOID been passed in 2004. Closest distance to Earth back in year 2069 on November 5 with MOID <0.05 au so 4179 Toutatis in that period can be classified as a potential hazardous asteroid. This conclusion support previous studies done by Sitarski (1998). The probability of the asteroid orbit intersecting Earth's orbit is essentially zero for at least the next six centuries. The collision in the future is considered to be very small. As a planet-crossing asteroid, the 4179 Toutatis is likely to be ejected from the Solar System by gravitational perturbation on a time scale of kilo of years, giving it a limited number to strike the Earth before disappearing forever. Another example of the phenomenon the asteroid ejected from the Solar System by gravitational force is 3552 Don Quixote (Siregar 2011). Previous studies by Benest et al.(1994) using the Lyapunov Characteristic Indicators show that the orbit of Toutatis has been found chaotic, through an interplay of the 3/1 resonance with Jupiter and close approaches with the inner planets; Venus, Earth and Mars. Hudson et al. (2003) showed that the rotation and angular velocity varies are caused by the density of matter are not distributed uniformly. Takahashi et al. (2012) founded that the distribution of Toutatis density is non-uniform with the strong concentration a 15% larger in head section (smaller lobe). Asteroid 4179 Toutatis is a clear sample of the complexity of motion that can be exhibited by purely gravitation bodies within the Solar System. More information should be retrieved from the exploration of the Chinese lunar probe Chang 2 which departed from the Sun-Earth Lagrange point L2 on April 15, 2012. The spacecraft is expected to make a flyby of 4179 Toutatis on December 13, 2012 (Lakdawalla 2012)

**Acknowledgments**. I thank the organizer of this conference. The orbital element data that used of this work supported by NASA (http://ssd.jplnasa.gov).